# Confocal microscopy through a multimode fiber using optical correlation

Damien Loterie,[1,*] Sebastianus A. Goorden,[2] Demetri Psaltis,[3] Christophe Moser[1]

[1] *Laboratory of Applied Photonics Devices, School of Engineering, École polytechnique fédérale de Lausanne, CH-1015 Lausanne, Switzerland*
[2] *Complex Photonic Systems, MESA+ Institute for Nanotechnology, University of Twente, P.O. Box 217, 7500 AE Enschede, The Netherlands*
[3] *Laboratory of Optics, School of Engineering, École polytechnique fédérale de Lausanne, CH-1015 Lausanne, Switzerland*
*Corresponding author: damien.loterie@epfl.ch

**We report on a method to obtain confocal imaging through multimode fibers using optical correlation. First, we measure the fiber's transmission matrix in a calibration step. This allows us to create focused spots at one end of the fiber by shaping the wavefront sent into it from the opposite end. These spots are scanned over a sample, and the light coming back from the sample via the fiber is optically correlated with the input pattern. We show that this achieves spatial selectivity in the detection. The technique is demonstrated on microbeads, a dried epithelial cell, and a cover glass.**

Multimode fibers are potentially interesting tools for minimally invasive endoscopy. Indeed, fibers can guide light with low loss over significant distances, while maintaining a very small outer diameter. At equal probe sizes, they also offer a much higher resolution than other fiber-based endoscopy solutions such as fiber bundles. The challenge is that modal scrambling distorts images as they propagate through multimode fibers: if an image is projected on one end facet of a multimode fiber, the opposite end shows a randomized version of this image.

However, modal scrambling can be compensated. In 1967 already, an experimental demonstration was made of the holographic reconstruction of a resolution target through a multimode fiber [1]. Later studies [2,3] explored other concepts for image restoration, but practical endoscopic imaging was difficult at the time due to technological limitations.

In recent years, the interest in this subject has been renewed due to developments in the field of wavefront control through distorting media [4]. Different groups [5–8] have now shown that the transmission characteristics of a fiber can be recorded on a computer during a calibration step, and that this data can be used in imaging to compensate for modal distortion effects. Various imaging modalities, such as reflection [8] and fluorescence [6,7], were subsequently demonstrated via a multimode fiber.

Several aspects of these systems are still actively under research, such as the bending stability. As a fiber is bent, its transmission characteristics can change, rendering a previous calibration obsolete. There is however a certain tolerance to bending [8,9], and a number of authors have noted that the fiber can be immobilized in a needle as a rigid ultrathin endoscope [6–8]. Such a system, capable of diffraction-limited resolution with an extremely thin probe (<300µm), could be used e.g. to assist in biopsies, since suspicious tissue could be analyzed microscopically in-situ prior to excision. Meanwhile, progress has also been made towards the dynamical compensation of bending [10–14], which may lead to fully flexible probes.

Another technical challenge is the optimal modulation of the wavefronts. Ideally, the targeted application (endoscopy) requires an accurate, high-speed and high-resolution modulator with good light efficiency. Currently available modulators offer a trade-off between those qualities: e.g. liquid crystal modulators typically achieve efficient and accurate phase modulation with speeds of only a few hundred Hz; digital micromirror devices reach kHz speeds but offer only binary amplitude modulation. There are strategies to overcome some of these limitations [6,15], and future modulators could improve substantially.

In this letter, we tackle the issue of image quality, and more precisely the suppression of background signals. When imaging inside a thick sample, light signals originating from any given image plane will always be superposed on an undesired contribution of light emanating from parts of the tissue outside this plane. This causes a blurring effect on the final image and decreases the overall contrast.

This problem can be solved in a number of ways. For example, two-photon fluorescence [13,16] or saturated excitation [17] have been proposed for imaging instead. As an alternative approach, we have demonstrated two computational methods to obtain confocal images via a multimode fiber [18]. Here, we improve upon these results and propose a way to obtain confocal images using optical correlation of the light signals returning from a multimode fiber. The main advantage of this optical implementation versus the digital approach reported before is that it is no longer necessary to record holograms during the imaging phase. This benefits the overall imaging speed (which is now limited only by the modulator), and the accuracy of the system: aberrations due to the non-flat surface of the SLM are automatically cancelled out when the field returns back to the SLM. This was not possible in the digital implementation, because the returning field was processed using a camera, i.e. a separate device.

The first step is to characterize modal scrambling. For this, we use a transmission matrix approach [5,6,8,19] as described in detail in our prior publications [18,20]. Briefly, we apply a series of linearly independent input patterns to one end of the fiber (Thorlabs M43L01, Ø105µm core, 0.22 NA) with a spatial light modulator (SLM, Fig. 1(a)), and we record the resulting output speckle patterns holographically on the opposite end (Fig. 1(b)). A complete set of such input-output measurements is called a transmission matrix. By inverting this matrix, it is possible to calculate which field needs to be shown on the SLM so that it creates a desired pattern on the opposite end, for example a spot. We refer to the side of the fiber with the modulator as the proximal side, and to the side where the sample is located as the distal side. During the calibration phase, the distal facet of the fiber is observed using an off-axis holographic acquisition system (Fig. 1(b)), but during imaging this hardware is no longer needed and only the sample should be present at the distal end (Fig. 1(c)).

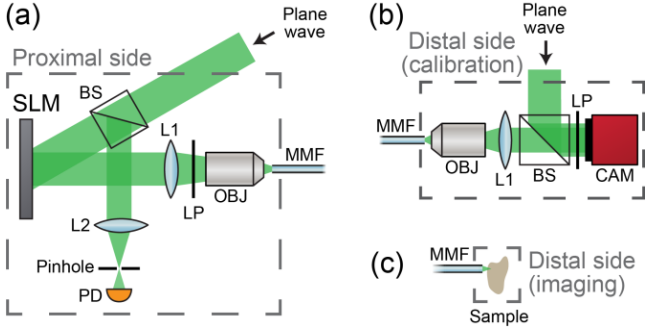

Fig. 1: Diagram of the experimental setup. BS: 50/50 beam splitter, SLM: HoloEye Pluto SLM, L1: f=250mm lens, L2: f=150mm lens, OBJ: Newport MV-40x objective, PD: Thorlabs PDA36A-EC photodiode, LP: linear polarizer, CAM: camera, MMF: multimode fiber. (a) Proximal side: an incoming plane wave is shaped by an SLM, and relayed to the input facet of a multimode fiber via a lens (L1) and a microscope objective (OBJ). The light signals returning from the fiber are relayed back onto the SLM, and focused via a lens (L2) through a pinhole. A photodiode (PD) records the resulting signal. (b) Distal side, calibration mode: an off-axis holographic system records the output fields from the fiber using a camera (CAM). (c) Distal side, imaging mode: a spot is scanned over the sample and the scattered/reflected light is collected back through the same fiber.

In the next step, we scan a spot over a rectangular grid inside the sample by appropriate modulation on the SLM. For each spot, the light scattered or reflected from the sample is collected back through the multimode fiber, and re-imaged onto the SLM (Fig. 1(a)). Here, this returning light is modulated by the illumination pattern shown on the SLM at that time. This is the first part of the correlation operation. If the returning light signal is originating from the same spot that is currently being illuminated by the SLM, then this returning signal retraces the same path as the illumination light back through the fiber. Therefore, the resulting wavefront at the SLM must be the phase conjugate of the illumination pattern. This can be deduced from the reversibility of wave equation. In literature, this principle is known as time reversal or phase conjugation [7,21]. If the returning light does not come back from the same spot that was illuminated (for example, it comes from a point in the background), then the resulting wavefront at the SLM will be decorrelated from the illumination pattern, due to the randomizing nature of modal scrambling.

After being modulated by the SLM, the returning signal is focused using a lens and then filtered by a pinhole in the Fourier plane of the lens. The pinhole extracts the zero-order term from the Fourier plane, i.e. the "average" of the incoming field. This completes the correlation operation: indeed, by multiplying a field with the conjugate of the pattern we want to extract from it, and then averaging out, we have carried out an optical equivalent [22–24] of the (non-normalized) correlation in mathematics, as shown in Eq. 1.

$$\rho = \sum_i x_i \cdot y_i^*  \quad (1)$$

The signal from the photodiode (PD in Fig. 1(a)) is proportional to $|\rho|^2$. A large amount of light returning from the illuminated spot results in a large photodiode signal. Light not originating from the illuminated spot is filtered away by the pinhole.

We calculated the point spread function (PSF) of this system using a numerical simulation. For this purpose, a synthetic transmission matrix was generated based on the theory of mode propagation in step-index multimode fibers. We assumed a straight fiber of 1m length, NA 0.22 and 105µm core. We then simulated each step of the scheme described before. The simulation takes into account the phase-only nature of the modulator, the modal scrambling of the fiber and the use of linear polarizers. It does not account for aberrations due to the optics, and it assumes paraxial propagation.

The longitudinal (xz) sections of the PSF are shown in Fig. 2 for various pinhole sizes: Fig. 2(a) is for a pinhole of 1 Airy unit, Fig. 2(b) is for 5 Airy units and Fig. 2(c) is the wide field case without a pinhole. Note that the Airy unit is used in confocal microscopy to denote the size of a diffraction-limited spot when imaged in the pinhole plane. This unit allows representing the pinhole diameter on a scale that is independent of the magnification of the optics used. In our case, the pinhole is not in an image plane but in the Fourier plane relative to the SLM. Therefore, the Airy unit is defined here by the Airy spot obtained as the Fourier image of the fiber core through OBJ, L1 and L2.

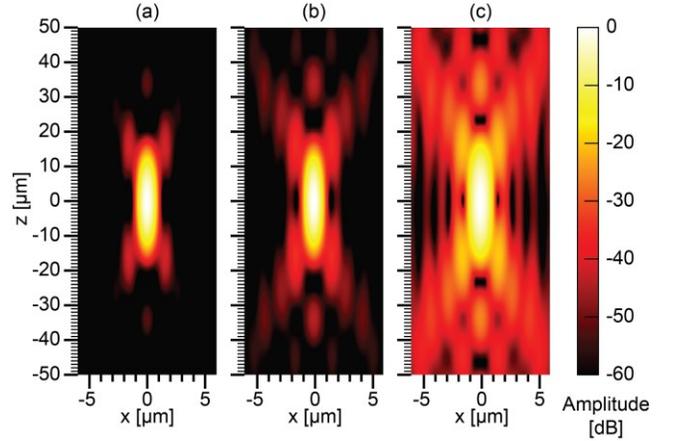

Fig. 2: Simulated point spread function for a pinhole size of (a) 1 Airy unit, (b) 5 Airy units, and (c) without pinhole. The images are rendered using a logarithmic color scale.

The PSF with a 1 Airy unit pinhole (Fig. 2(a)) is roughly proportional in magnitude to the square of the PSF without pinhole (Fig. 2(c)). The total response integrated over each transverse plane (xy-slice) is comparable in every plane of the PSF without pinhole, whereas it decreases quickly as we move away from the focus when using a pinhole (sectioning effect). The lateral full width at half maximum (FWHM) resolutions are 0.95µm, 0.96µm and 1.3µm respectively for Figs. 2(a), 2(b) and 2(c). The axial FWHM resolutions are 15.4µm, 15.9µm and 21µm respectively.

We experimentally verified the validity of the correlation confocal procedure on multiple samples. For comparison purposes, we used similar samples as in our previous study using computational processing [18]. We made images "without" pinhole (Figs. 3(a), 3(c) and 3(e)) and with a 30μm pinhole, which is approximately 1 Airy unit in our implementation (Figs. 3(b), 3(d) and 3(f)). The images "without" pinhole actually use a large 2mm pinhole, because otherwise stray light signals (e.g. an unmodulated portion of light from the SLM) also reach the detector; these signals are not related to the sample and would make a comparison inaccurate.

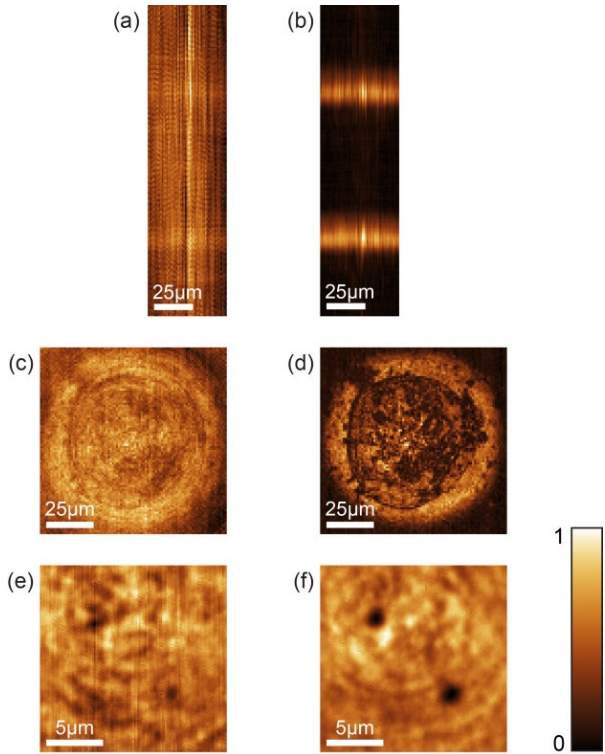

Fig. 3: Experimental results (a,c,e) with a 2mm pinhole and (b,d,f) with a 30μm pinhole. (a-b) Depth-scan of a 150μm cover glass. (c-d) Lateral scan of a human epithelial cell on the surface of a cover glass. (e-f) 1μm polystyrene beads on the surface of a cover glass. Each image is normalized between 0 and 1, where 0 is the minimum photodiode voltage in this image and 1 the maximum voltage.

As shown in Fig. 3(b), the optical correlation method clearly resolves the reflective interfaces of a cover glass. This is not possible without pinhole (Fig. 3(a)). In the case of an epithelial cell (Fig. 3(d)) or polystyrene beads (Fig. 3(f)), the correlation method dramatically increases the obtained contrast versus non-filtered images (Figs. 3(c) and 3(e)). To quantify the axial resolution, we calculated the average FWHM of the interface in Fig. 3(b), which is 14.9μm. The lateral FWHM resolution was estimated from a lateral scan of a 100nm nanoparticle on a cover glass, and is approximately 1.3μm. We believe the difference with the simulation is due to the low signal levels when measuring small nanoparticles with the current NA. The point-scanning rate is limited by our spatial light modulator at 20Hz.

In our experiment, we have used a fiber with NA 0.22 to limit the size of the transmission matrix (940Mb of computer memory using double-precision complex numbers). This facilitates processing with commonly available computer resources. We note however that the scheme is adaptable to fibers with arbitrary NA, as well as other usage cases such as scattering media.

In conclusion, we reported on an all-optical method to obtain confocal images through a multimode fiber. This method uses only a spatial light modulator to improve imaging contrast and give a sectioning capability. These results could be relevant in future applications such as multimode fiber endoscopy in thick biological tissues.

**Funding.** Swiss National Science Foundation (200021_160113/1, project MuxWave).